\documentstyle[11pt,aaspp4,flushrt,psfig]{article}
\def\Jpb{Jy~beam$^{-1}$}
\def\eg{{e.g.}}
\def\ie{{i.e.}}

\def\etal{{et al.}}

\def\aL4m{$\alpha_{74M}^{1.5}$}
\def\aPL{$\Delta\alpha_{0.3}^{1.5}$}
\def\aPC{$\Delta\alpha_{0.3}^{5}$}

\newcommand{\ex}[1]{\mbox{$\times 10^{#1}$}}

\newcommand{\sn}[1]{\mbox{s$^{#1}$}}

\newcommand{\uv}{\mbox{\em u-v}}
\newcommand{\as}{\mbox{\char125}}

\def\Ra  #1 #2 #3.#4{\mbox{$ #1^h  #2^m #3.\!\!^s#4$}}
\def\dec #1 #2 #3.#4{\mbox{$ #1\arcdeg #2' #3.\!\!^s#4$}}

\begin{document}

\title{The Radio Spectral Index of the Crab Nebula}
\author{M. F. Bietenholz}
\affil{Department of Physics and Astronomy, York University, 
North York, M3J~1P3, Ontario, Canada}

\author{N. Kassim}
\affil{Code 7213, Remote Sensing Division, Naval Research Laboratory, Washington, DC 20375-5351}

\author{D. A. Frail and R. A. Perley}
\affil{National Radio Astronomy Observatory, Socorro, New Mexico, 87801, USA}

\author{W. C. Erickson}
\affil{University of Maryland, College Park, Maryland, 20742 USA}

\author{A. R. Hajian}
\affil{United States Naval Observatory, Washington, DC 20392-5100, USA}

\clearpage
\begin{abstract}

\noindent We present the results of a new, comprehensive investigation
of the radio spectral index of the Crab Nebula supernova remnant.  New
data at 74~MHz are combined with data at 327~MHz, 1.5~GHz and 5~GHz.
In contrast to previous claims, little spatial variation in the
spectral index is seen. In particular, between 327 MHz and 5 GHz we
see no evidence of spectral steepening near the edge of the nebula,
the ``jet'' or the ionized filaments. The rms limits on any spectral
index variations in these regions amount to no more than 0.01. We
believe that earlier reports of large steepening were the result of
correlator bias and image registration problems. An elongated feature
was detected 1\arcmin\ northwest of the pulsar which may be a
continuation of the well-known wisp-like structures seen closer to the
center of the nebula.  At 74~MHz, we see for the first time evidence
of free-free absorption by the thermal material in the Crab Nebula's
filaments. Apart from some possible renewed acceleration occurring in
the wisps, the dominant accelerator of relativistic electrons in the
Crab Nebula is the pulsar itself.

\end{abstract}

\keywords{ISM: Individual (Crab Nebula) --- Radio Continuum: ISM --- supernova 
remnants}

\section{Introduction}

The Crab Nebula supernova remnant (SNR) is one of the strongest radio
sources in the sky and has long been an object of intense scrutiny by
radio astronomers. Nevertheless, a number of controversies with
important physical implications continue to surround the details of
its radio continuum spectrum. While there is general agreement that
the Crab's integrated spectrum has a slope $\alpha=-0.27\pm{0.04}$
$\footnote{We define the radio spectral index, $\alpha$, so that
$S_\nu \propto \nu^\alpha, S_\nu$ being the flux density at frequency
$\nu$}$ from 10 MHz to the synchrotron break frequency near 10$^4$ GHz
(Strom \& Greidanus 1992, Kovalenko, Pynzar' \& Udal'tsov 1994), there
is considerable dispute over the details of its spatial distribution
(Wilson 1972, Swinbank 1980, Velusamy, Roshi \& Venugopal 1992,
Bietenholz \& Kronberg 1992, hereafter BK92). Since the spectral index
serves as a useful measure of the energy distribution of the
relativistic electrons which generate the synchrotron emission, claims
of {\em spatial variations}\/ in spectral index across the nebula
therefore must be interpreted in terms of differences in the
underlying physical processes which influence the energy of the
particles.

The following are the three main areas where small variations in the
radio spectral index, $\alpha$, have been reported: In the first
instance, BK92 report spectral index variations in the vicinity of the
pulsar, similar to the wisps found in the optical (Scargle 1969,
Hester \etal\ 1995).  Although they can be seen as weak
arcsecond-width features in the total intensity images, these
structures are most prominent in the spectral index images. They have
a flatter spectral index than the nebula, suggesting that they are
regions of renewed particle acceleration. These spectral index ridges
are located in the same region but are not positionally coincident
with the optical wisps. They have the same concave arc-like appearance
as the optical wisps but are somewhat longer.  These ridges were interpreted
as manifestations of the wind shock in the Crab by BK92, and Kronberg
\etal\ (1993) modeled them as such with a tilted ring or torus geometry.

Secondly, Velusamy \etal\ (1992) have claimed that the spectrum of the
filaments in the body of the nebula is different than that of the
intra-filament plasma, and that there is a trend of steepening filament
spectral index as a function of radius in the nebula. BK92, however,
reported no such variations.  If true, these spectral variations must
be reconciled with interpretations of the physical nature of the
filaments and the source of their radio emission (Reynolds 1988).
This would imply a different, or at least an additional acceleration
process for the relativistic electrons in the filaments.

Thirdly, several groups have reported a steepening of the spectral
index near the edge of the nebula.  (Velusamy \& Sarma 1977, Agafonov
1987) These measurements, however, are difficult to make, and other
groups report no steepening of $\alpha$ towards the edges of the
nebula (Wilson 1972, Swinbank 1980, Trushkin 1986, BK92).  If such a
steepening is real, then it could be interpreted as evidence for a
{\em shell}\/ of emission around the Crab, analogous to the steeper
spectrum shell-type emission typically observed from classical
shell-type SNRs (\eg\ Cas-A). The question of whether or not such a
shell exists superposed on the plerion is extremely important, since
if it does, this would successfully bring to an end the search for the
classical radio shell exterior to the presently observed Crab SNR.
The lack of such a shell has led previous workers to important
conclusions regarding the energy of the supernova explosion, the
efficiency of its blast wave in accelerating relativistic particles,
and the nature of the circumstellar environment surrounding the SN
precursor star (Nomoto 1985; Nomoto 1987; Frail \etal\ 1995, F95
hereafter). If the existence of a shell were confirmed, all
conclusions drawn from those negative searches would be in error.

In order to tackle these controversies, we bring together in this
paper a comprehensive body of observational material, and present here
important new sub-arcminute resolution 74~MHz observations.  This
radio database, encompassing an almost two decade range in frequency,
allows us to conduct the most sensitive study of the magnitude and
spatial distribution of the Crab's radio spectral index conducted to
date. We then use this unique body of evidence to address the
observational controversies outlined above, and to address the
physical implications of these findings.

\section {Observations and Data Reduction}

\begin{figure}
\begin{center}
{\bf Figures}\\
\vspace{3cm}
\leavevmode\psfig{figure=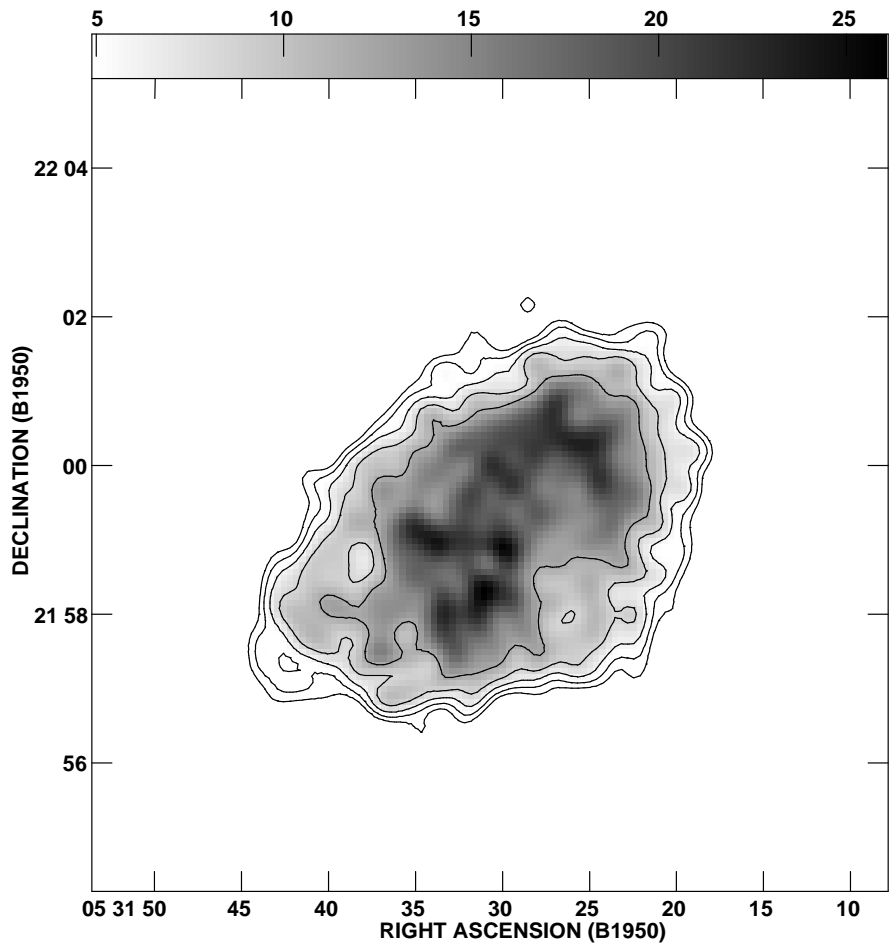,width=12cm}
\end{center}
\caption{The 74~MHz image of the Crab, with the pulsar
subtracted.  The FWHM of the restoring beam was $25\as$.  The peak
flux density is 26~\Jpb, and the noise level is 0.5~\Jpb.  The contours
are drawn at 11.3, 16, 22.6, 32, and 45.3\% of the peak flux density.}
\label{74img}
\end{figure}

We used several sets of radio observations as described in Table~1.
All data were taken at the NRAO VLA\footnote{The NRAO Very Large Array
is a facility of the National Science Foundation operated under
cooperative agreement by Associated Universities, Inc.}.  These
include 74~MHz observations which currently utilize only eight VLA
antennas. The 74~MHz data comprised snapshot observations from three
A, two B, and one C array configuration spanning several years in
order to obtain sufficient \uv\ coverage. The non-standard procedures
employed to reduce these data are described in Kassim \etal\
1993.  The image obtained at 74~MHz is shown in Fig.~\ref{74img}.

Several problems which have not been addressed fully in the past
complicate the determination of the spectral index from interferometer
data, and especially from the VLA.

Firstly, the images at different frequencies need to be registered
accurately.  Most radio images have had the antenna phases improved by
self-calibration (see, for example Pearson \& Readhead, 1984; or
Perley, 1989) Once this has been done, the absolute position
information is degraded.  To make a spectral index map, however, the
two maps must be registered accurately, and so, a way of recovering
the absolute position is needed.  Typically this problem can be
ignored, for if the starting model is reasonable the shift in position
is small, and registering maps by the nominal positions is sufficient.
Nonetheless, shifts of up to 1/4 of a beamwidth can occur.  In the
case of a completely resolved source, a shift of only 1/10 of a beam
can change the map amplitude at a particular pixel position by up to
25\%, which can cause substantial spurious spectral index variations,
especially near bright gradients such as those near the edges of SNRs.

Images of the Crab are complicated, and registering them to less than
1/10 beamwidths is difficult in the presence of noise.  One means of
accomplishing this is by using the MIRIAD program IMDIFF (Dickel
\etal\ 1988) which uses the maximum entropy method of Tan and Gull
(1985) to find the translation required to make the first image
resemble the second one as closely as possible in a least squares
sense (see Bietenholz \etal, 1991, for an application of this method
to radio images of the Crab). The other approach is to self-calibrate
the data in phase using a model of known positional offset, which will
serve to bring the positions to the reference frame defined by the
model (i.e. using the image at one frequency to phase calibrate the
data at another frequency).

A much less tractable problem, peculiar to the VLA, is the correlator
non-linearity (Kulkarni \& Heiles 1980).  The correction needed to
compensate for this is known as the quantization or van Vleck
correction, and it is {\em not}\/ done at the VLA.  To our knowledge,
none of the previous VLA results have addressed this problem.  The
output of the digital correlator employed at the VLA is a known
function of the true analog correlation factor.  For small correlated
flux densities, this function is very nearly linear, and this linear
approximation is what is used at the VLA (D'Addario 1989).  However,
for objects as strong as the Crab, the relation between the digital
correlation and the true analog correlation is no longer linear, and
the derived correlated flux densities will be in error.  This error
increases with the observed flux density.  It affects the real and
imaginary part of the correlation coefficient independently, and since
the uncalibrated phase, \ie\ that of the correlated flux density, is
random, it will cause random phase and amplitude errors.  The
observed amplitudes, however, will always be lower than the true ones,
so the resulting maps will be {\em biased}.  For our data, at 5~GHz,
these errors can be as high as 4\% in amplitude.

Unfortunately, the full information needed to correct for this effect
is not as yet preserved at the VLA, so it is difficult to accurately
recover the true correlation after the fact.  This is less likely to
affect our images at 74~MHz because the correlator non-linearity is
less severe at this frequency, and our signal-to-noise ratio is considerably
lower.  For the data at other bands, we investigated the effects of
this problem by making two sets of spectral index maps: First, the
``standard'' version which is made in the straightforward way, with
the maps being registered by IMDIFF.~ Secondly we made ``careful''
spectral index maps, in which we deconvolved using only data at \uv\
distances $> 1.6 $K$\lambda$, where the correlated flux density is low enough
that the error in neglecting the correlator bias is minor.  The
registration was achieved by also self-calibrating in phase using a
common model.  We describe the production of these images in more
detail below.

Whether any significant bias is introduced into the maps by the
correlator non-linearity is difficult to estimate.  The process of
deconvolution routinely recovers the flux density in the center of the
\uv~plane starting from an estimate of zero, so one might think that
it should have no problem recovering a few percent of missing flux density.
On the other hand the deconvolution process is completely free to
adjust the flux density in the center of the \uv~plane, whereas in regions
where there are measurements, deconvolution will attempt to fit the
biased measurements.  We found that the apparent noise levels in the
maps deconvolved by our ``careful'' process was lower than the
``standard'' ones.  This suggests that there {\em is}\/ discernible
noise introduced by the missing correction, since one doesn't expect
the rms to decrease when one reduces the amount of data.  This would
suggest that the bias might also be significant.

We will now describe more fully the production of the ``careful''
spectral index images: Because the measurements near the center of the
\uv~plane contain the most correlated flux, they are most severely
affected by the correlator non-linearity.  We made maps without using
the data compromised in this way.  We chose to use a \uv~distance
cutoff rather than a flux density cutoff because its effect on the images is
easier to understand.  The resulting spectral index images will
contain little information on angular scales larger than $2.7'$, \ie\
half the size of the nebula.  At 5~GHz, which is the worst case, the
correlated flux density at this \uv~distance is $\sim20$~Jy and the
remaining correlator non-linearity errors will be $<0.05$\%.

Since the Crab Nebula is considerably larger than the largest scale of
$2.7'$ which we are sampling, we will recover the large scale
structure by deconvolving with a default (support) in our maximum
entropy deconvolutions (AIPS task VTESS; see Cornwell \& Evans 1985,
Cornwell 1988).  We will use the same default (appropriately scaled)
for all our observing frequencies, which is justified because the
spectral index is quite uniform over the nebula (Wilson 1972, BK92).
We will concentrate on mapping the {\em difference}\/ to the Nebula's
average spectral index rather than its absolute spectral index, which
can be more reliably determined using other methods: Kovalenko \etal\
(1994) have recently determined it to be $\alpha = -0.27\pm{0.04}$.
Using the same default map will bias our maps to look similar, and
thus any spatial variation in spectral index will be {\em required}\/
by the data (this is the ``forced'' VTESS technique of Anderson \&
Rudnick 1993).

The default map was produced as follows: At each frequency (except at
74~MHz) we mapped the data without any \uv~restrictions. The resulting
restored maps (\ie\ convolved with an appropriate ``clean'' beam and
with the residuals added) were corrected for the primary beam pattern,
and for the expansion of the Crab (0.13\% per year, Bietenholz \etal\
1991).  They were then convolved to a common resolution of $16\as
\times 16\as$, and flux-scaled to a common frequency using the Baars
\etal\ (1977) spectral index of $\alpha = -0.3$.  The recent Kovalenko
\etal\ (1994) value of $\alpha = -0.27$ is consistent with the older
Baars \etal\ value to within the uncertainty, so using the latter should not
affect our results. These maps were then averaged, and the result was
used as the default map in all the subsequent deconvolution processes.
For each deconvolution, it was again scaled back to the flux density
and epoch appropriate for that frequency.

Furthermore, in order to assure accurately registered maps, this same
(scaled and expanded) default image was used to self-calibrate, in
phase only, the \uv~data prior to deconvolution.  This process has the
side effect of further biasing the maps to be as similar as is allowed
by the data.  In other words it will minimize any spatial variations
in the resulting spectral index map.  However, our dynamic range is
high, and we will show that this bias is negligible, and that our
``careful'' spectral index maps can be used with confidence.

Once self-calibrated in this way, the \uv~data were transformed into
the image plane, using only data at \uv~distances greater than
1.6~K$\lambda$, and then deconvolved using the AIPS task VTESS and
the appropriately scaled default map.

The resulting restored maps, were convolved to a common beamsize,
contracted to a epoch 1987.4 and then spectral index maps were made in
the usual way.  The uncertainties were derived from empty regions of
the maps near the nebula (especially at 5~GHz, the noise in a map
corrected for primary beam pattern increases with distance from the
phase center).  These noise estimates were consistent with the
differences between 4625 and 4885~MHz maps. At 74~MHz the
pulsar is quite strong, and we first subtracted it from the fully
calibrated \uv~data.  Furthermore, the 74~MHz observations have much
lower signal to noise and fewer antennas, and the correlator
non-linearity is less severe, so we did not make a ``careful''
spectral index map involving this frequency.

\section{Results}

\begin{figure}
\begin{center}
\leavevmode
\psfig{figure=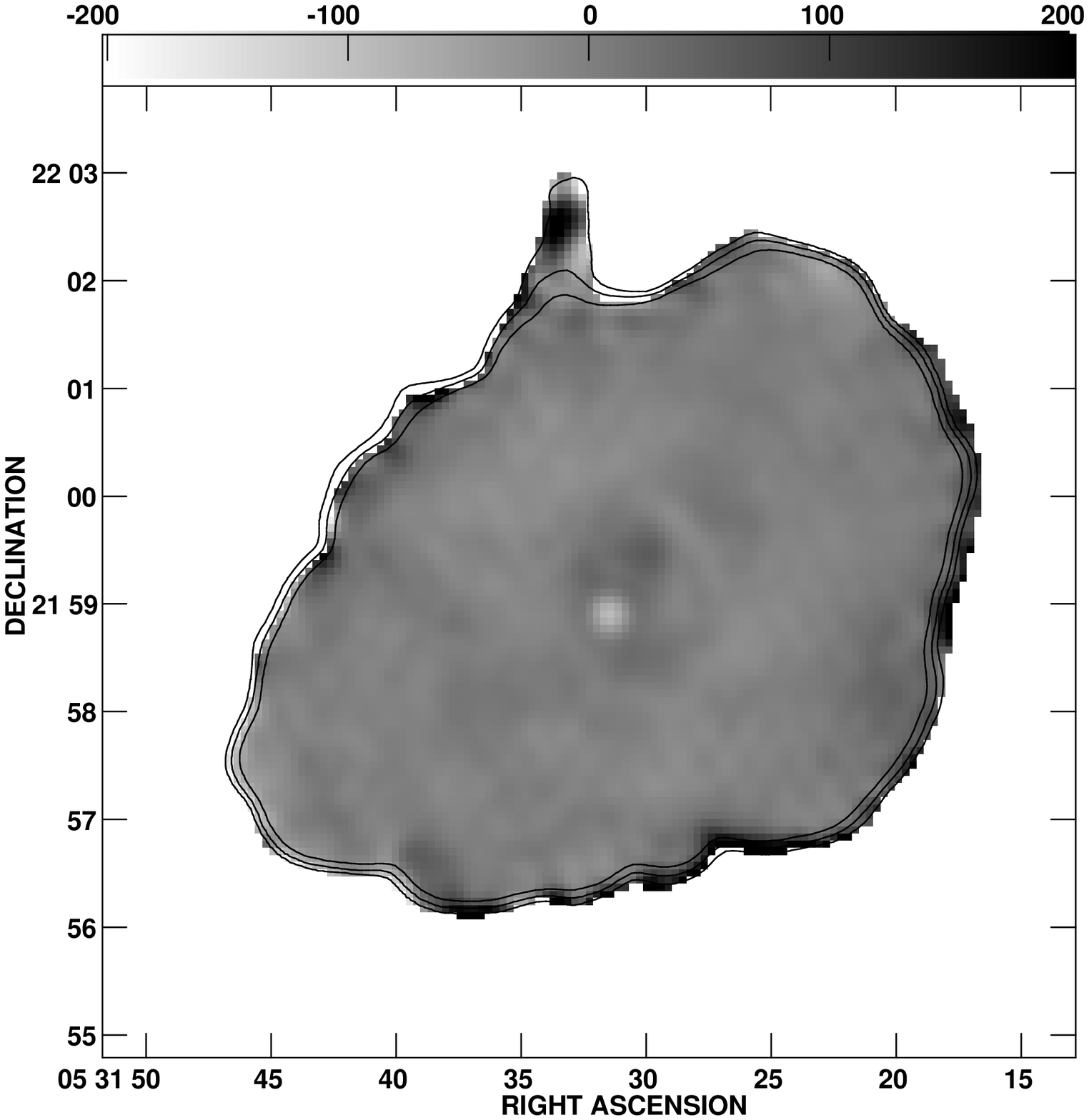,width=7.7cm}
\psfig{figure=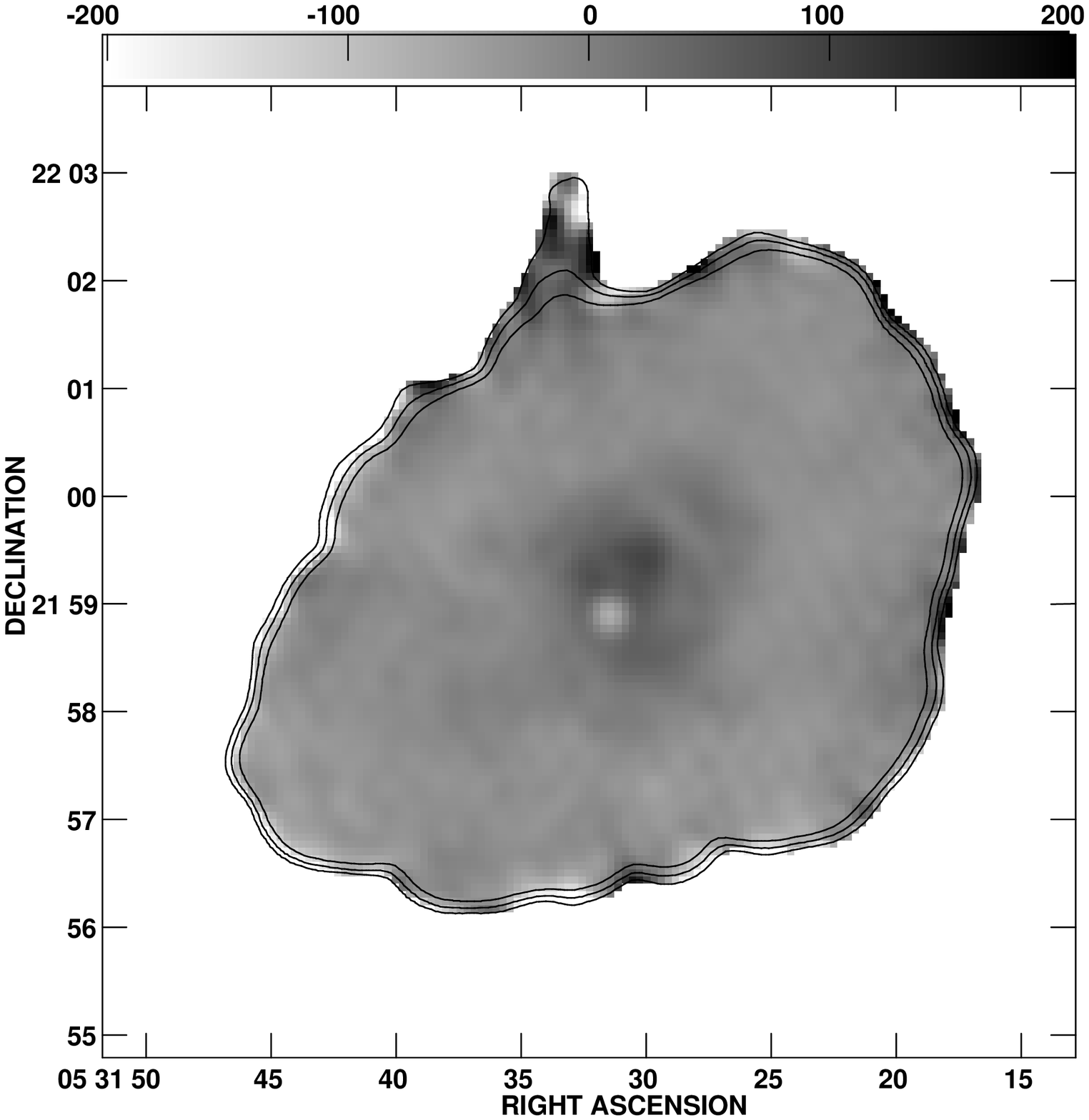,width=7.7cm}\\
$(a)$ \hspace{7cm} $(b)$
\end{center}
\caption{The relative radio spectral index between 327~MHz
and 1.5~GHz (\aPL) in units of 0.001.  These maps show the difference
of the spectral index to the mean spectral index of the nebula.  The
FWHM of restoring beam is $16\as$.  The contours are the 0.5, 1, and
2\% contours in total intensity of the 327~MHz image. $(a)$ The
``careful'' version, made by deconvolving with a default and using
only data at \uv~distances $> 1.6$K$\lambda$.  It contains little
information on angular scales $> 2.7'$, but is not corrupted by
correlator non-linearity.  $(b)$ The ``standard'' version, which was
made without a default, and contains information on all angular scales
but may be slightly biased.}
\label{spxPL}
\end{figure}

\begin{figure}
\begin{center}
\leavevmode
\psfig{figure=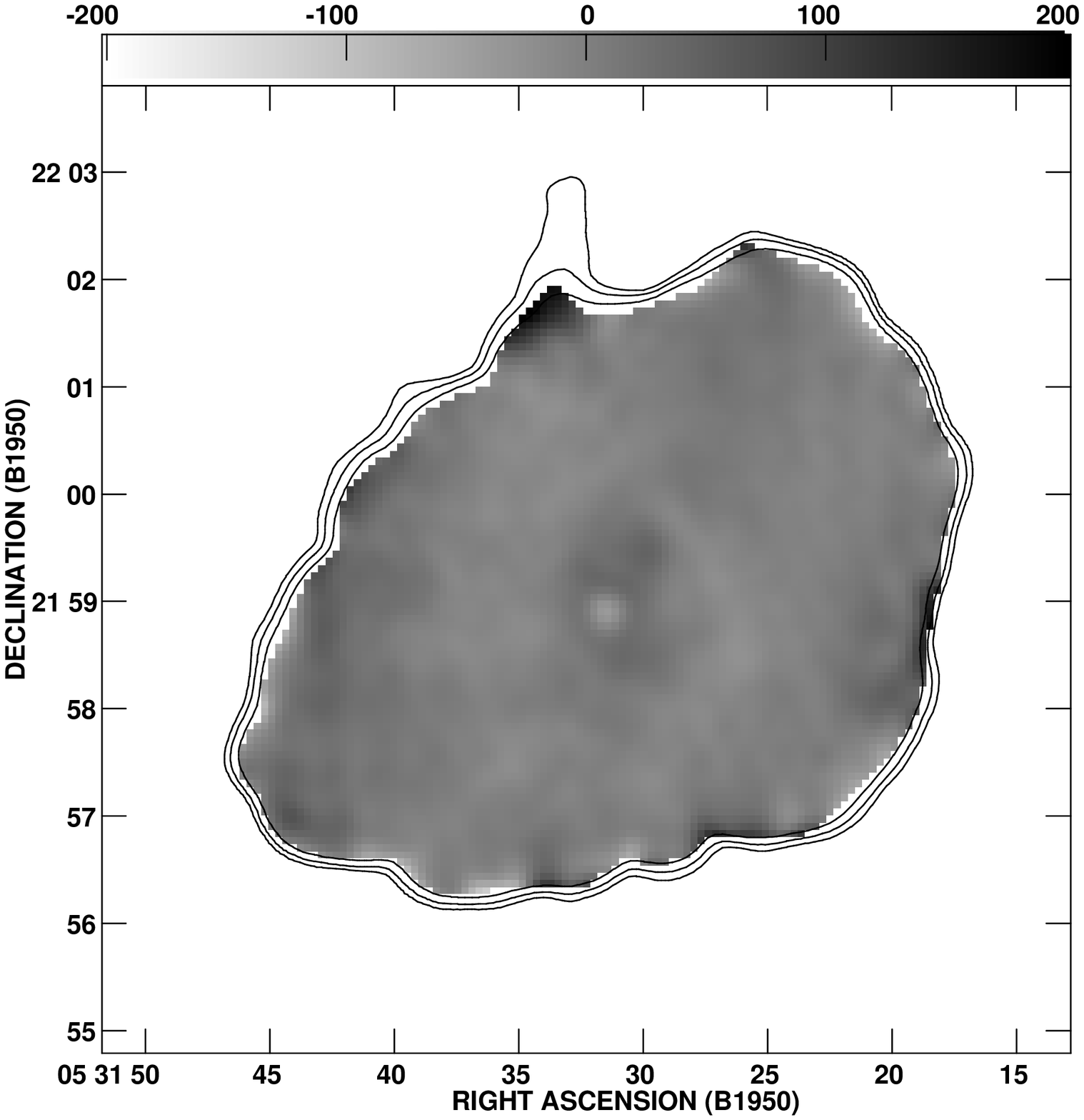,width=7.7cm}
\psfig{figure=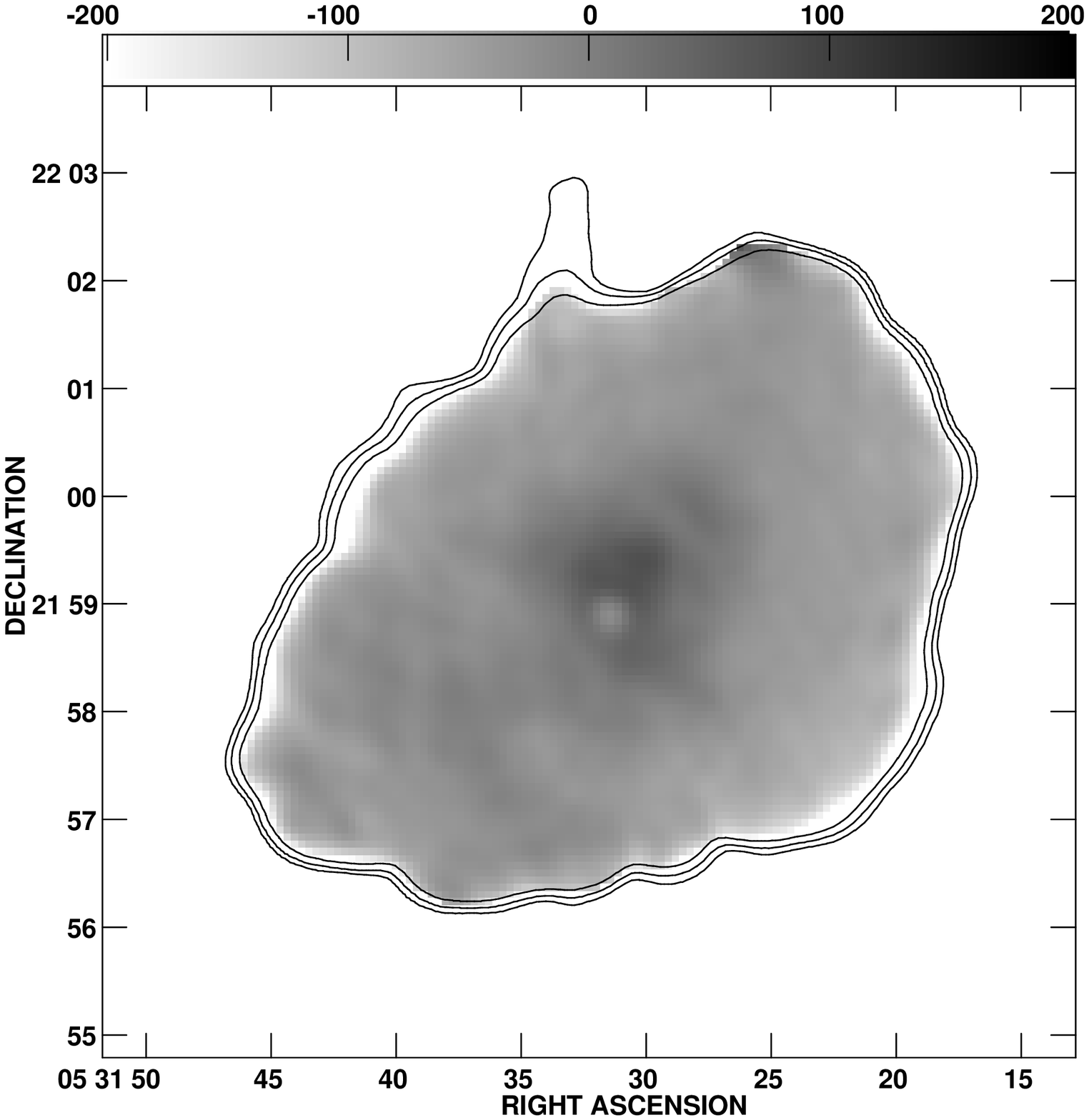,width=7.7cm} \\
$(a)$ \hspace{7cm} $(b)$
\end{center}
\caption{The relative radio spectral index between 327~MHz
and 5~GHz (\aPC) in units of 0.001.  These maps show the difference of
the spectral index to the mean spectral index of the nebula.  The FWHM
of restoring beam is $16\as$.  The contours are the 0.5, 1, and 2\%
contours in total intensity of the 327~MHz image. $(a)$ The
``careful'' version and $(b)$ is the ``standard'' version (see Fig.~2
and text, \S2).}
\label{spxPC}
\end{figure}

\begin{figure}
\begin{center}
\leavevmode\psfig{figure=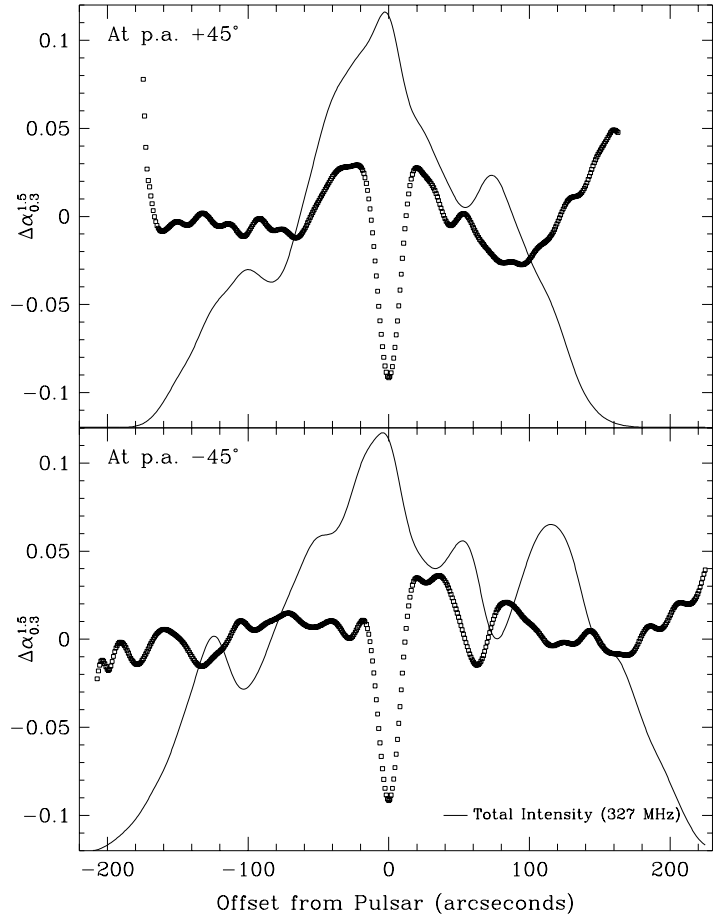,width=10cm}
\end{center}
\caption{Two profiles of the \aPL\ spectral index map
(``careful'' version).  The profiles go through the pulsar position at
the indicated position angles ($-45$\arcdeg and +45\arcdeg).  Squares
indicate \aPL, while the thin line indicates the total intensity at
327~MHz, included for comparison.}
\label{profil}
\end{figure}

The relative spectral index between 327~MHz and 1.5~GHz (\aPL) is
shown in Fig.~\ref{spxPL} while Fig.~\ref{spxPC} shows the relative
spectral index between 327~MHz and 4.9~GHz (\aPC); both at a
resolution of $16\as \times 16\as$. These are spectral indexes
relative to the mean spectral index of the nebula ($-0.27$): we are
not sensitive to the total flux densities, so our integrated spectral
indexes are not constrained by our data.  Our reduction process will
make them very close to the value of $-0.3$ which we assumed for the
default maps.  In addition, Fig.~\ref{profil} shows two profiles through the
\aPL\ spectral index map.  The profiles go through the pulsar position
at position angles $-45$\arcdeg\ and +45\arcdeg.

We have included both the ``careful'' and the ``standard'' maps in
order to assess the significance of the bias (see \S{2}) and the
degree of success we have had in treating it. Both maps reveal that
the spectral index of the nebula is quite uniform.  The rms over the
whole nebula of \aPL, for example, is only 0.01 in $\alpha$ when one
excludes the region near the edge of the nebula where the
signal-to-noise is low.  The differences between the ``careful'' and
the ``standard'' maps are negligible in the case of \aPL: the rms of
the difference in $\alpha$ is 0.008. There is no evidence of spectral
steepening of any significant magnitude near the edges of the nebula.
An ``apparent'' steepening is visible at some locations, near the
nebular boundaries. However, these occur at brightness levels less than
1\% of the peak brightness of the Crab Nebula (at 327 MHz), where the
uncertainty in $\Delta\alpha$ is large. 

The pulsar is the strongest feature in these spectral index maps.  It
appears as a point source of steeper spectral index visible in the
\aPL\ and \aPC\ maps.  It shows up as a steepening of only 0.07 on the
\aPC\ map.  This translates into a pulsar flux density of 1.9~Jy at 327~MHz,
which is consistent with the value of 1.3~Jy, given by Lorimer \etal\
(1995), since pulsar fluxes are known to be quite variable.  Since the
spectral index at the pulsar did not change noticeably between the
``careful'' and ``standard'' maps, any bias introduced by our
``careful'' procedure is small.  All the maps also show a slight
flattening in a region of about $1'$ to $2'$ in size surrounding the
pulsar, and extending slightly farther to the NW.  The magnitude of
this effect ranges from about $\Delta \alpha = 0.05$ on the
``standard'' \aPC\ map to $\Delta \alpha = 0.02$ on the ``careful''
\aPC\ and the \aPL\ maps.  It is notably larger on the ``standard''
\aPC\ map than on the ``careful'' one.  Since the angular scales
involved are those which could be biased by the correlator
non-linearity, which is worst at 5~GHz, we suspect that the estimate
of $\Delta \alpha = 0.02$ is more reliable.  This effect is gradual,
and remaining systematic sources of error are probably at about this
level.  We note that the correlator non-linearity would tend to reduce
the observed flux density especially at higher frequencies and at large
angular scales, which could cause a slight spurious spectral
index steepening to appear towards the edges of the nebula in VLA
images.

We emphasize that our conclusions above are largely independent of
whether we examine the ``careful'' or the ``standard'' spectral index
maps; this suggests that the bias introduced by the correlator
non-linearity is in fact small.  Our relatively uniform spectral index
images suggest that the visibilities at two frequencies should
differ only by a scale factor.  We examined the gridded visibility
data, and they show the expected linear relationship between 327 and
1.5~GHz at \uv-distances $> 1.6$K$\lambda$.  The formal scatter is
about $2\sigma$, which is compatible with the small spectral index
differences found (since the above process accounted for neither residual
calibration errors nor the primary beam pattern, we expect the scatter
about a linear relationship to be somewhat larger than 1 even in the
case of a completely uniform spectral index).

The ``wisps'' which were visible in the higher resolution spectral
index map of BK82 are barely discernible at this resolution. This is
simply because our beam area is about 70 times that of BK82.  Another
feature {\em is}\/ consistently visible on all the \aPC\ and \aPL\
maps: it is a lane of steeper spectral index about $1'$ to the NW of
the pulsar.  This feature has an \aPL\ about 0.03 steeper than the
surrounding region. 

The spectral index of the Crab Nebula's ``jet'' is insignificantly
steeper than that of the body of the nebula. Formally, \aPL\ of the
jet is $+0.05 \pm 0.07$ relative to the rest of the nebula.  We
conclude that the jet seems to have the same spectral index as the
rest of the nebula.  This is consistent with the star trail model of
Cox, Gull \& Greene (1991), which explains the jet as the star trail
of the Crab's progenitor.  The star trail has cooled into a hollow
tube, and the lower pressure inside the tube causes the relativistic
plasma from the nebula to expand more rapidly there.

\begin{figure}
\begin{center}
\leavevmode\psfig{figure=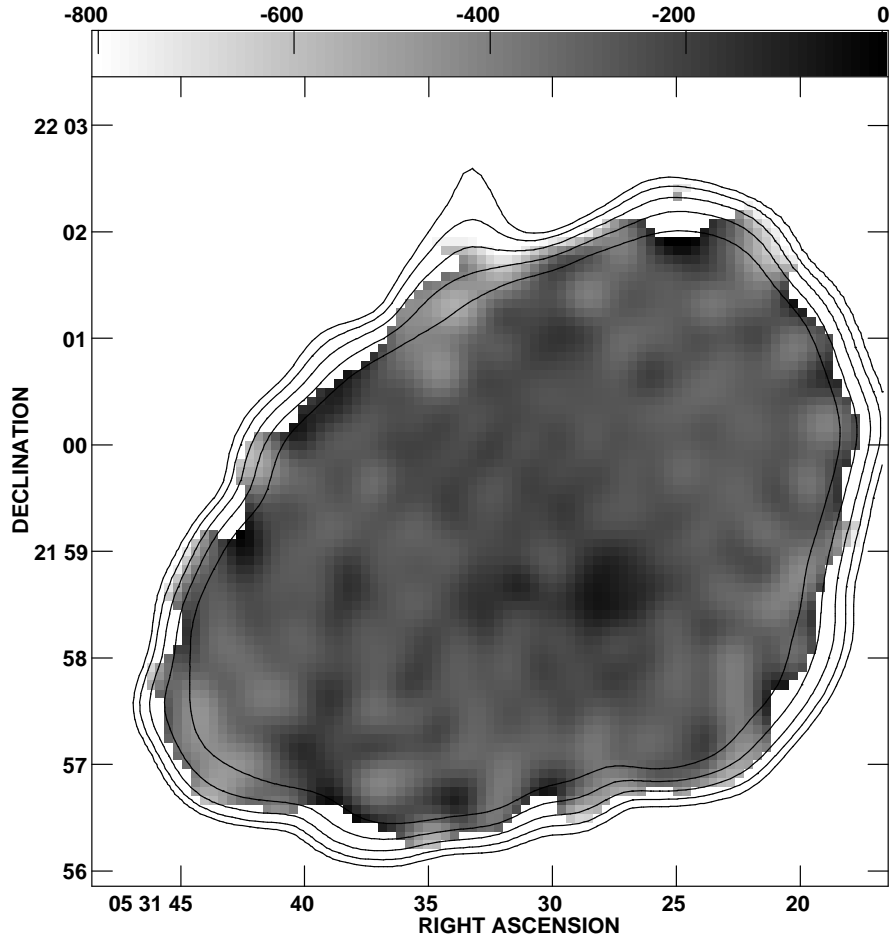,width=12cm}
\end{center}
\caption{The radio spectral index between 74~MHz and
1.5~GHz (\aL4m) in units of 0.001.  The size of the restoring beam was
$27\as \times 25\as$ elongated at position angle $-33\arcdeg$.  The
contours are in total intensity at 1.5~GHz, drawn at 0.5,1,2, 4 and
8\% of the peak flux density.}
\label{spxL4m}
\end{figure}

Fig.~\ref{spxL4m} shows the map of the spectral index between 74~MHz and 1.5~GHz
(\aL4m), this time at the slightly coarser resolution of $27\as \times
25\as$.  Because of the lower signal-to-noise of the 74~MHz map and
because the correlator non-linearity is small at this lowest
frequency, we expect the errors caused by not correcting for it will
be less than the noise. The pulsar is, of course, not visible in the
74~MHz map, nor the \aL4m\ map because it has already been removed
from the 74~MHz data.  However, some significant structure is visible
in this map.  The most distinct feature on this map is at RA = \Ra 05
31 27.7, DEC = \dec 21 58 42.0 and has spectral index $\sim 0.3$ {\em
flatter}\/ than the rest of the nebula.  It is most likely a low
frequency {\em absorption}\/ feature because no corresponding feature
is visible at frequencies above 327~MHz.  Its reality was tested by
deconvolving the 74~MHz data using an appropriately scaled, translated
and convolved version of the 1.5~GHz map as a default --- once again
biasing the maps to be as similar as is allowed by the data, and it
did not diminish in strength. Furthermore, since the 74~MHz data is
derived from a number of observing runs, and the feature in question
is not confined to only one of them, it is almost certainly neither
an instrumental or a deconvolution effect.

\section{Discussion}

We see no steepening of the spectral index near the edge of the
nebula, confirming the result of BK92.  We find no support for the
hypothesis proposed by Velusamy \& Sarma (1977) that the emission near
the edge of the nebula comes from a supernova shell which is
superimposed on the plerionic emission from the rest of the nebula.
On our very sensitive spectral index maps between 74~MHz and 5 GHz we
find that the mean spectral index near the edge of the
nebula\footnote{The region 'near the edge' was defined as the region
between the 2\% and 20\% contours of the 5~GHz total intensity map, on
which the peak flux density was 5.2~\Jpb.  The quoted uncertainties
are the rms of $\alpha$ per beamwidth over those regions.} is $-0.01
\pm 0.04$ {\em flatter}\/ than that in the interior of the nebula.
The spectral index between 74~MHz and 1.5~GHz (\aL4m) shows no
steepening to the edge of the nebula either, so the spectrum near the
edge maintains its power-law form at least down to 74~MHz. The
uniformity of the spectral index throughout the nebula suggests that
all the synchrotron-\-emitting electrons have the same origin, implying
that they are all accelerated by the pulsar.

Velusamy \etal\ (1992) claim that the spectral index of the {\em
filaments}\/ ($\alpha_f$) as opposed to that of the smooth continuum
emission, steepens gradually towards the edge of the nebula, with
$\alpha_f$ increasing smoothly by 0.3 from the center to the edge of the
nebula.  Such a steepening would be apparent in our data even though
we do not separate the filament emission from the continuum:
Velusamy \etal\ (1992) find that the filaments account for 20\% of the
total flux density at 327~MHz, and since their filling factor is $<1/2$
we estimate that their contribution to the surface brightness at
filament locations should be $> 40\%$.  This would cause the 
$\alpha$ of the combined emission from the continuum and filaments
to change by at least 0.1 which is considerably larger than
any spectral index change we observe.

A very slight and gradual steepening of the spectrum {\em is}\/
observed as one moves away from the pulsar.  The magnitude of this
effect is about 0.02 in $\alpha$. While the reality of this feature
was questioned in \S{3}, we may expect some steepening to occur, since
the radio-emitting electrons are moving outward and losing energy by
synchrotron radiation.  However, unlike in much longer-lived
extragalactic sources, where synchrotron aging can dominate (Carilli
\etal\ 1991), the magnitude of this effect may be too small to
account for this slight observed steepening of 0.02 in $\alpha$: If we
assume an (average) nebular field of 5\ex{-5}~G and that the electrons
radiate all their power at $\nu_m = 1.22 (B/$Gauss$) \gamma^2$~MHz we
can estimate the change in spectral index of only $<0.01$ over the lifetime
of the nebula.

The visible Crab Nebula is an expanding bubble of thermal filaments
and relativistic gas.  It must, at some level, interact with the
exterior medium into which it expands.  This interaction would
probably take the form of a shell.  If it is expanding into the ISM,
this shell would presumably be similar to other, shell-like supernova
remnants.  At 327~MHz, F95 detected no extended shell emission {\em
exterior}\/ to the nebula, and thus no evidence for a fast hydrogen
envelope around the Crab.  If there is no fast envelope around the
visible nebula, however, then the interface between the nebula and the
exterior medium is at the observed edge of the nebula.  This interface
presumably involves a shock which could accelerate particles.  The
emission from this interaction would thus be superposed on the visible
nebula.  Shock acceleration typically produces radio spectra steeper
($\alpha \sim -0.5$) than that of the pulsar-accelerated interior of
the nebula, so if there were a significant quantity of
shock-accelerated electrons near the outer boundary, a steepening of
the radio spectral index would be expected.  We find no such
steepening or any ``rim'' around the outer edges of the nebula.

Taking the limit of the change in $\alpha$ near the edge to be $0.03$,
we find that the maximum residual surface brightness with $\alpha =
-0.5$ near edge of the nebula is $< 0.15$~\Jpb\ at 327~MHz.  This
would represent $<60$~Jy if spread evenly over the nebula, which would
entail a fraction of less than 2\% of the total radio luminosity of
the Crab (integrated from $10^7$ to $10^{11}$ Hz); and even less if
the $\alpha = -0.5$ emission displayed a shell morphology.
This would correspond to a minimum relativistic particle energy in the
putative shell of less than few $\times 10^{48}$~ergs (using standard
assumptions from Pacholczyk 1970).  While shell-type supernova remnants
typically convert between 1\% and 10\% of the blast energy into
shock-accelerated synchrotron-emitting particles (Duric \etal\ 1995),
the Crab's blast wave is expected to have converted a far smaller
fraction of its energy into particles for two reasons: Firstly, the
Crab is not yet in the Sedov phase of its evolution, and calculations
show the efficiency of particle acceleration is very low in the
free-expansion phase (Drury Markiewicz \& V\"{o}lk 1989, Markiewicz,
Drury \& V\"{o}lk 1990). In any case, since the shell is observed to
be accelerating (Bietenholz \etal\ 1991) it cannot have converted a
great deal of its original energy into relativistic particles.
Secondly, the Crab supernova may well have been an anomalously
low-energy event (Nomoto 1985, Nomoto 1987, Pols 1994, F95).
In the case that the visible Crab is expanding into a halo of
pre-SN wind or ejecta, as hypothesized by Hester \etal\ (1996), we
would also expect $<1\%$ of the energy of the expanding bubble 
to have been converted in accelerated particles since the relative
velocity of the synchrotron bubble is lower relative to the exterior medium. 
Again, we would then not expect to see synchrotron emission from the
shock-accelerated electrons at the edge of the synchrotron bubble.

The lane of steeper spectral index mentioned in \S2\ is about $1'$ to
the NW of the pulsar.  This corresponds in position to the moving
feature found in the radio by BK92, and to the moving `dark lane'
visible in the optical (Oort \& Walraven 1956, Scargle 1969).  This
feature is moving outward, and in the radio, it seems to have a sharp
outward edge (BK92).  Since our 327~MHz data was taken several years
{\em after}\/ the data at 1.5 and 5~GHz, we would expect such a moving
feature to generate spurious steeper spectral index feature.  We can
roughly estimate the feature's proper motion to be about $0.4 \pm
0.2$~asec~yr$^{-1}$, representing a speed of $\sim 4000$~kms$^{-1}$.
This is marginally consistent with the optical proper motion of $0.7
\pm 0.1$~asec~yr$^{-1}$ (Scargle 1969).  Possibly, the feature is
slowing down, since BK92 find a marginally slower speed
(0.6~asec~yr$^{-1}$) in 1985.  This is consistent with the feature being
embedded in the large-scale flow in the nebula.  Higher resolution
observations would be needed to accurately determine the speed of the
feature.

\section{Low-Frequency Absorption}

As discussed above, the significant feature SW of the pulsar on the
\aL4m\ spectral index map has an index ~0.3 flatter than the rest of
the nebula, and suggests there is low-frequency absorption towards
this location. This feature has not been seen in prior spectral index
maps, and it is the availability of a sub-arcminute resolution image
at 74~MHz that has enabled its detection. Comparison with the
three-dimensional [OIII] data of Lawrence \etal\ (1995) indicates that
it corresponds to the location of the brightest optical knot with
negative radial velocity. This places the knot on the near side of the
nebula and thus supports our hypothesis that the flat spectral index
feature is due to free-free absorption by thermal electrons in the
line emitting filaments.

\begin{figure}
\begin{center}
\leavevmode\psfig{figure=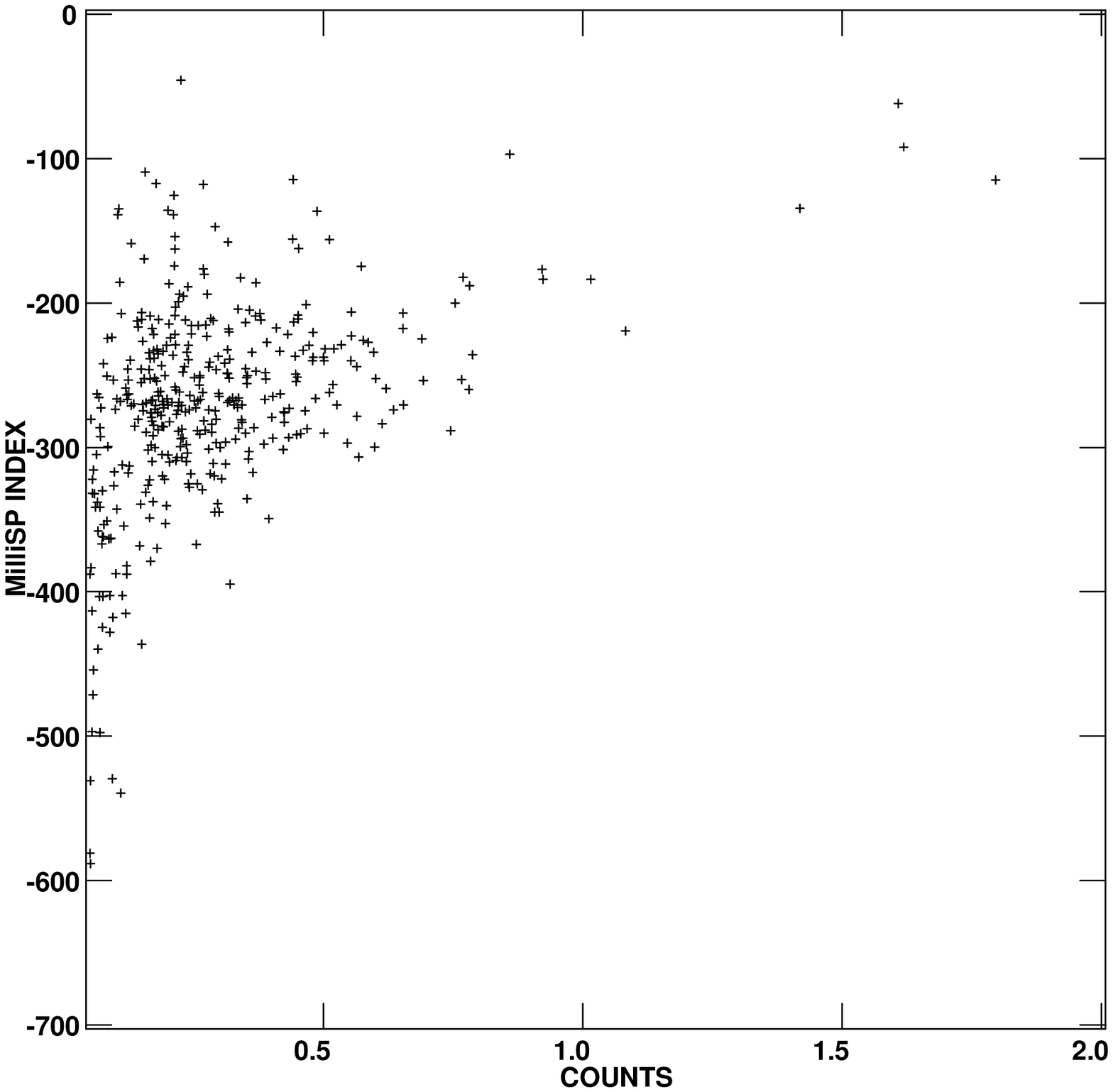,width=10cm,angle=-90}
\end{center}
\figcaption{Plot of intensity of [OIII] $\lambda 5007$\AA\
emission with negative radial velocity, i.e.\ that originating from the
side of the nebula towards us vs.\ the 74~MHz to 1.5~GHz spectral
index (\aL4m).  Optical data kindly supplied by S. Lawrence (Lawrence
\etal\ 1995).  A correlation is evident, indicating that the thermal
material in the filaments which emits OIII line emission is causing
free-free absorption of the radio continuum.}
\label{OIII}
\end{figure}

The rms variations of the spectral index relative to the mean are $2.4
\sigma$ when excluding the aforementioned feature, indicating that
there {\em is}\/ real structure in the spectral index.  A plot of the
intensity of the [OIII] emission from the front of the nebula, \ie\
that with negative radial velocity, against the \aL4m\ spectral index
(Fig.~\ref{OIII}) shows that there is a weak correlation between the two, in the
sense that the \aL4m\ is flatter, \ie\ the low-frequency absorption is
higher at the locations where there is more [OIII] emission.  This is
what would be expected if the thermal material responsible for the
[OIII] emission is causing free-free absorption at 74~MHz.

Assuming an electron temperature of $T = 10^4$K for the optically
emitting thermal filaments, our measure of the free-free optical depth
constrains the emission measure to be $\ga 800$~cm$^{-6}$pc, the
inequality arising because not all of the synchrotron emission will
originate from behind the absorbing feature.  Assuming a line-of-sight
depth similar to the feature's width, we calculate that the electron
density in the filaments $n_e \ga 200$ cm$^{-3}$, is compatible with
the values expected for the Crab's filaments (Davidson \& Fesen 1985).
This is also compatible with densities derived from the depolarization
of the Crab between 5~GHz and 1.5~GHz (Bietenholz \& Kronberg 1991).
In fact, a prominent depolarization feature can be seen at this
location.

\section{Summary}

We have presented new, high-resolution 74~MHz observations of the Crab
Nebula.  Combining these with a large body of observational material
at other radio frequencies up to 5~GHz, we have made sensitive radio
spectral index maps of the Crab Nebula.  We have carefully registered
our maps at different frequencies, and we have taken steps to
eliminate the effects of the correlator non-linearity, which will tend
to contaminate data of bright objects taken at the VLA.  In agreement
with some earlier work, we find that the radio spectral index,
$\alpha$, of the Crab is quite uniform over the nebula at frequencies
above 300~MHz.

We also find that:
\begin{enumerate}

\item{There is {\em no}\/ steepening of the spectral index near the edge of
the nebula.  There is thus no evidence for any steeper spectrum {\em
shell}\/ emission around the Crab.  Such emission is expected to be faint
since the Crab is still in the free-expansion phase.}

\item{Any gradual steepening towards the edge of the nebula is small,
being on the order of 0.02 in $\alpha$.  This is much smaller than has
been claimed in the past.  Synchrotron radiation losses probably
produce an effect somewhat smaller than this, although it is of about this
order of magnitude, and hence they may be responsible.}

\item{The spectral index of the filaments is no different from
that of the body of the nebula for frequencies above 300~MHz, which
suggests that the origin of the relativistic electrons in the
filaments is the same as that in the body of the nebula.}

\item{We confirm the existence of a moving radio feature in the NW
quadrant of the nebula.  It corresponds in position to the dark
lane noted by Oort and Walraven (1956; see also Scargle 1969).
It appears to be moving at $\sim 4000$~km~\sn{-1}.}

\item{At frequencies below 300~MHz, we see evidence of the free-free
absorption caused by the thermal material in the optical line emitting
filaments.  Estimates of the electron density in the line emitting
filaments derived from the amount of free-free absorption we measure are
compatible with other estimates derived from the optical spectra
of the filaments.}

\end{enumerate}

\acknowledgements We would like to thank S. Lawrence for kindly
supplying us with his [OIII] data.  Research at York University was
partly supported by NSERC.  NRAO is operated under license by
Associated Universities, Inc., under cooperative agreement with NSF.
Basic research in radio astronomy at the Naval Research Laboratory is
supported by the Office of Naval Research.  Finally, we thank the
referee, Dr.\ Richard Strom, for his useful comments.

\clearpage

\begin{table}
\begin{center}
\begin{minipage}{6.0in} % So the footnotes appear 
\small
\centerline{TABLE 1}
\vspace{0.05in}
\centerline{Details of the observing sessions}
\vspace{0.05in}
\begin{tabular}{ l c  l l}
\hline\hline
Frequencies& epoch  & VLA arrays used & Reference \\
~~(MHz)     \\
\hline
4885, 4625 & 1987   & BCD             & Bietenholz \& Kronberg 1991 \\
1515, 1410 & 1987   & ABCD            & Bietenholz \& Kronberg 1991 \\
327        & 1992   & BC              & Frail \etal\ 1995 \\
74         & 1993   & ABC             & Kassim \etal\ 1993 \\
\hline
\vspace{2in}
~
\end{tabular}
\end{minipage}
\end{center}
\end{table}

\begin{references}
\reference{} Agafonov, M. I., \etal\ 1987, \sovast, 64, 60
\reference{} Anderson, M. C., \& Rudnick, L. 1993, \apj, 408, 514
\reference{} Baars, J. W. M., Genzel, R.,  Pauliny-Toth, I. I. K., \& Witzel, A. 1977, \aap,
61, 99 
\reference{} Bietenholz, M. F., \& Kronberg, P. P. 1991, \apj, 368, 231 
\reference{} Bietenholz, M. F., \& Kronberg, P. P. 1992, \apj, 393, 206 (BK92) 
\reference{} Bietenholz, M. F., Kronberg, P. P.,Hogg, D. E., \& Wilson, A. S. 1991, \apjl, 373, L59 
\reference{} Carilli, C. L., Perley, R. A., Dreher, J., W., \& Leahy, J. P. 1991, \apj, 338, 554
\reference{} Cornwell, T. J. 1988, \aap, 202, 316 
\reference{} Cornwell, T. J., \& Evans, K. F., 1985, \aap, 143, 77 
\reference{} Cox, C. I., Gull, S. F., \& Greene, D. A. 1991, \mnras, 250, 750 
\reference{} D'Addario, L. R. 1989, in ASP Conf.\ Proc.\ 6, Synthesis Imaging in Radio
Astronomy, ed.\ R. A. Perley, R. R.  Schwab, \& A. H. Bridle (San Francisco: ASP), 59 
\reference{} Davidson, K., \& Fesen, R. A. 1985, \araa, 23, 119 
\reference{} Dickel, J. R., Sault, R., Arendt, R. G., Matsui, Y., \& Korista, K. T. 1988, \apj, 330, 254 
\reference{} Drury, L. O'C., Markiewicz, W. J., \& V\"{o}lk, H. J. 1989, \aap, 225, 179
\reference{} Duric, N., Gordon, S. M., Goss, W. M., Viallefond, F., \& Lacey, C. 1995, \apj, 445, 173 
\reference{} Frail, D. A., Kassim, N. E., Cornwell, T. J., \& Goss, W. M. 1995 \apjl, 454, L129 (F95)
\reference{} Hester, J. J. \etal. 1995, \apj, 448, 240 
\reference{} Hester, J. J. \etal. 1996, \apj, 456, 225
\reference{} Kovalenko, A. V., Pynzar', A. V., \& Udal'tsov, V. A. 1994, Astr.~Reports, 38, 110 
\reference{} Kassim, N. E., Perley, R. A., Dwarakanath, K. S., \& Erickson, W. C. 1993, \aj, 106, 2218
\reference{} Kronberg, P. P., Lesch, H., Ortiz, P. F., \& Bietenholz, M. F. 1993,
\apj, 416, 251
\reference{} Kulkarni, S. R., \& Heiles, C. 1980, \aj, 85, 1413
\reference{} Lawrence, S. S., MacAlpine, G. M., Uomoto, A., Woodgate,
B. E., Brown, L. W., Oliversen, R. J., Lowenthal, J. D., \& Liu, C. 1995,
\aj, 109, 2653
\reference{} Lorimer, D. R., Yates, J. A., Lyne, A. G., \& Gould, D. M. 1995, \mnras, 273 411
%\reference{} Manchester, R. N., \& Taylor, J. H. 1977 Pulsars
\reference{} Markiewicz, W. J., Drury, L. O'C., \& V\"{o}lk, H. J. 1990, \aap, 487, 502
\reference{} Nomoto, K. 1985, in The Crab Nebula and Related 
Supernova Remnants, eds.\ M. Kafatos \& R. B. C. Henry (Cambridge: 
Cambridge University Press), 97 
\reference{} Nomoto, K. 1987, in The Origin and Evolution of
Neutron Stars, IAU Symp.~125, ed.\ D. J. Helfland \& J.-H. Huang
(Dordrecht: Reidel), 281 
\reference{} Oort, J. H. \& Walraven,
Th. 1956, Bull.\ Astron.\ Inst.\ Netherlands, 12, 285.  
\reference{} Pacholczyk, A. G. 1970 Radio Astrophysics (San Francisco:
W. H. Freeman) `
\reference{} Perley, R. A. 1989, in ASP Conf.\ Proc.\
6, Synthesis Imaging in Radio Astronomy, ed.\ R. A. Perley,
R. R. Schwab, \& A. H. Bridle (San Francisco: ASP), 287.  
\reference{} Pearson, T. J., \& Readhead, A. C. S. 1984, \araa, 22, 97
\reference{} Pols, O. R. 1994, \aap, 290, 119 
\reference{} Reynolds, S. P. 1988, \apj, 327, 853 
\reference{} Scargle, J. D. 1969, \apj, 156, 401
\reference{} Strom, R. G., \& Greidanus, H. 1992, Nature, 358, 654
\reference{} Swinbank, E. 1980, \mnras, 193, 451 
\reference{} Tan, S. M., \& Gull, S. F. 1985, \mnras, 216, 949 
\reference{} Trushkin, S. A. 1986, Soviet Astron.\ Lett., 12, 81 
\reference{} Velusamy, T., \& Sarma, N. V. G. 1977, \mnras, 181, 455 
\reference{} Velusamy, T., Roshi D., \& Venugopal V. R. 1992 \mnras,
255, 210 
\reference{} Wilson, A. S. 1972 \mnras, 157, 229
\end{references}
\end{document}